          [2001/04/25 1.1 (PWD)]
\documentclass[rnote]{aa}

\def\p1{\phantom{0}}
\def\x1{\phantom{00}}
\def\y1{\phantom{000}}
\def\z1{\phantom{0000}}

\usepackage{graphicx,color}
\usepackage{natbib}
\bibpunct[, ]{(}{)}{;}{a}{}{,}
\begin{document}

\titlerunning{Goldilocks planet Gliese 581g}

\title{Habitability of the Goldilocks planet Gliese 581g:\\
       Results from geodynamic models}
\author{W. von Bloh
\inst{1}
\and
M. Cuntz
\inst{2}
\and
S. Franck
\inst{1}
\and
C. Bounama
\inst{1}
}
\offprints{W. von Bloh}
\institute{Potsdam Institute for Climate Impact Research, P.O. Box 60 12 03, 14412 Potsdam, Germany\\
\email{bloh@pik-potsdam.de}
\and
Department of Physics, University of Texas at Arlington, Box 19059, Arlington, TX 76019, USA}
\date{Received <date> / Accepted <date>}
\abstract{}
{In 2010, detailed observations have been published that seem to indicate another super-Earth planet
in the system of Gliese 581 located in the midst of the stellar climatological habitable zone.  The
mass of the planet, known as Gl~581g, has been estimated to be between 3.1 and 4.3~$M_\oplus$.  In this
study, we investigate the habitability of Gl~581g based on a previously used concept that explores
its long-term possibility of photosynthetic biomass production, which has already been used to
gauge the principal possibility of life regarding the super-Earths Gl~581c and Gl~581d.}  
{A thermal evolution model for super-Earths is used to calculate the sources and sinks of
atmospheric carbon dioxide.  The habitable zone is determined by the limits of photosynthetic
biological productivity on the planetary surface.  Models with different ratios
of land / ocean coverage are pursued.}
{The maximum time span for habitable conditions is attained for water worlds at a position of
about $0.14\pm 0.015$~AU, which deviates by just a few percent (depending on the adopted stellar
luminosity) from the actual position of Gl~581g, an estimate that does however not reflect
systematic uncertainties inherent in our model.  Therefore, in the framework of our model an almost
perfect Goldilock position is realized.  The existence of habitability is found to critically
depend on the relative planetary continental area, lending a considerable advantage
to the possibility of life if Gl~581g's ocean coverage is relatively high.}
{Our results are a further step toward identifying the possibility of life beyond the Solar System,
especially concerning super-Earth planets, which appear to be more abundant than previously surmised.}


\maketitle


\section{Introduction}

The nearby M-type dwarf Gliese 581 is an important target of previous and ongoing planetary search
efforts, and so far four, but possibly up to six planets around this star have been discovered.  It is
noteworthy that until recently the possible existence of Earth-type and super-Earth planets in
extrasolar systems has been highly speculative \citep[e.g.,][]{butler06}.
\cite{bonfils05} reported the detection of a Neptune-size planet around Gl~581 (M3~V).
Gl~581 is at a distance of 6.26 pc; it has a mass of about $0.31~M_{\odot}$ and a luminosity of
$0.013 \pm 0.002~L_{\odot}$ (see below).

Later on, \cite{udry07} announced the detection of two so-called ``super-Earth" planets
in this system, Gl~581c and Gl~581d, with minimum masses of
5.06 and 8.3~$M_\oplus$ and with semi-major axes of 0.073 and 0.25~AU, respectively.
In 2009, Gl~581e was discovered, another super-Earth planet \citep{mayor09}.
Meanwhile the position of Gl~581d has been re-identified as 0.22~AU. Finally, in 2010,
\cite{vogt10} reported a still controversial discovery.  The authors argue in favor of the
existence of two additional super-Earth planets, namely Gl~581f and Gl~581g.  Note that
Gl~581g is of particular interest to the study of planetary science and astrobiology because it is
positioned in the midst of the stellar climatological habitable zone.  Its mass is estimated
to be between 3.1 and 4.3~$M_\oplus$, noting that the putative upper mass limit is partially
constrained by the assumption that none of the planetary orbital eccentricities exceeds 0.2.
The distance of Gl~581g is estimated to be $0.14601 \pm 0.00014$~AU, and its eccentricity is
almost identical to zero.

According to \cite{valencia06}, super-Earths are rocky planets from one to about ten Earth masses
with a chemical and mineral composition akin to that of Earth.  In the following, we adopt the
hypothesis that this is indeed the case, and consider for Gl~581g our model previously developed
for the Earth, using scaling laws if appropriate.  This approach has already been applied to
Gl~581c and Gl~581d \citep{vonbloh07}, as well as to simulations of hypothetical super-Earth planets
in various other systems, i.e., 47~UMa and 55~Cnc \citep{cuntz03,franck03,vonbloh03}, as well as
systems undergoing red giant branch evolution \citep{vonbloh09}.

The main question is whether Gl~581g, if existing, offers the principal possibility of life, i.e.,
whether it lies within the stellar habitable zone (HZ), as already argued by \cite{vogt10}.
Typically, stellar HZs are defined as regions around the central star where the physical
conditions are favourable for liquid water to exist at the planet's surface for a period of
time sufficient for biological evolution to occur.  \cite{kasting93} calculated the HZ boundaries
for the luminosity and effective temperature of the present Sun as $R_{\mathrm{in}} = 0.82$ AU
and $R_{\mathrm{out}} = 1.62$ AU.  They defined the HZ of an Earth-like planet as the region
where liquid water is present at the surface.

According to this definition, the inner boundary
of the HZ is determined by the loss of water via photolysis and hydrogen escape.  The outer
boundary of the HZ is determined by the condensation of CO$_2$ crystals out of the atmosphere
that attenuate the incident sunlight by Mie scattering.  The critical CO$_2$ partial
pressure for the onset of this effect is about 5 to 6 bar.  However,
the cooling effect of CO$_2$ clouds has been challenged by \cite{forget97}.
CO$_2$ clouds have the additional effect of reflecting the outgoing thermal
radiation back to the surface.  The precise inner and outer limits of the
climatic habitable zone are still unknown due to the limitations of the
existing climate models.  Recently, \cite{heller10} extended the concept of the HZ
by including constraints arising from tidal processes due to the planet's spin orientation
and rate.  Effects caused by tilt erosion, tidal heating, and tidal equilibrium rotation seem
to be especially important for potentially habitable planets (as Gl~581d and Gl~581g)
around low-mass stars.  For limitations of the planetary habitability of M-type stars see,
e.g., \cite{tarter07}, \cite{guinan09}, and \cite{segura11}.

The luminosity and age of the central star play important roles in the
manifestation of habitability.  The luminosity of Gl~581 can be obtained by
(1) photometry \citep{bonfils05,udry07}, and (2) the application of the
mass--radius relationship \citep{ribas06} together with the spectroscopically
determined stellar effective temperature of $T_e = 3480$~K
\citep{bean06}.  Both methods yield $L=0.013 \pm 0.002~L_\odot$.
\cite{selsis07} estimated the stellar age as at least 7~Gyr based
on the non-detection of Gl~581's X-ray flux considering the
sensitivity limit of ROSAT \citep{schmitt95,voges00}.

In the following, we adopt a definition of the HZ previously used by
\cite{franck00a,franck00b}.  Here habitability
at all times does not just depend on the parameters of the central star, but
also on the properties of the planet.  In particular, habitability is linked
to the photosynthetic activity of the planet, which in turn depends on the
planetary atmospheric CO$_2$ concentration together with the presence of
liquid water, and is thus strongly influenced by the planetary dynamics.
We call this definition the photosynthesis-sustaining habitable zone, pHZ.  
In principle, this leads to additional spatial {\it and
temporal} limitations of habitability because the pHZ (defined for a specific
type of planet) becomes narrower with time owing to the persistent decrease of
the planetary atmospheric CO$_2$ concentration.


\section{Habitability of super-Earth planets}

\subsection{Photosynthesis-sustaining habitable zone (pHZ)}

The climatic habitable zone at a given time for a star with luminosity $L$ and
effective temperature $T_e$ different from the Sun can be calculated 
following \cite{underwood03} based on previous work by \cite{kasting93} as
\begin{equation}
R_{\mathrm{in}} = \left(\frac{L}{L_\odot\cdot S_{\mathrm{in}}(T_e)}\right)^{\frac{1}{2}}~,~~R_{\mathrm{out}}
                = \left(\frac{L}{L_\odot\cdot S_{\mathrm{out}}(T_e)}\right)^{\frac{1}{2}}
\label{hz_jones}
\end{equation}
with $S_{\mathrm{in}}(T_e)$ and $S_{\mathrm{out}}(T_e)$ described
as second order polynomials.

To assess the habitability of a terrestrial planet, an Earth-system model
is applied to calculate the evolution of the temperature and 
atmospheric CO$_2$ concentration.  On Earth, the carbonate--silicate cycle
is the crucial element for a long-term homeostasis under increasing solar
luminosity.  On geological time-scales, the deeper parts of the Earth are
considerable sinks and sources of carbon.

\begin{figure}
\centering
\resizebox{0.7 \hsize}{!}{\includegraphics[width=1cm]{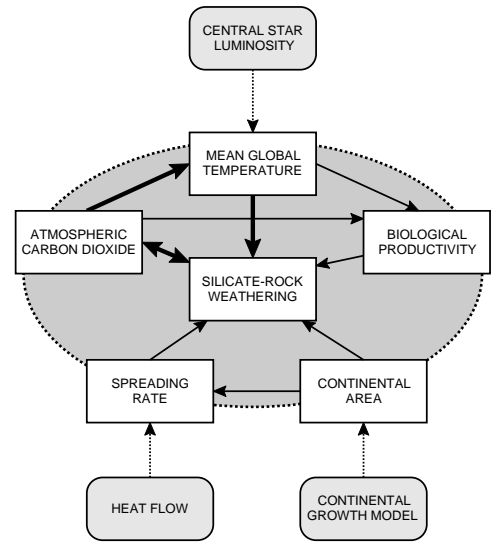}}
\caption{Earth system box model.  The arrows indicate the different forcing
and feedback mechanisms.  The bold arrows indicate negative feedback operating
towards climate stabilization.
} \label{block}
\end{figure}

Our numerical model previously applied to Gl~581c and Gl~581d \citep{vonbloh07}
couples the stellar luminosity $L$, the silicate--rock
weathering rate $F_{\mathrm{wr}}$ and the global energy balance to obtain
estimates of the partial pressure of atmospheric carbon dioxide
$P_{\mathrm{CO}_2}$, the mean global surface
temperature $T_{\mathrm{surf}}$, and the biological productivity $\Pi$ as
a function of time $t$ (Fig.~\ref{block}).  The main point is the
persistent balance between the CO$_2$ sink in the atmosphere--ocean system and
the metamorphic (plate tectonic) sources.  This is expressed through the
dimensionless quantities  
\begin{equation}
f_{\mathrm{wr}}(t) \cdot f_A(t) = f_{\mathrm{sr}}(t),
\label{gfr}
\end{equation}
where $f_{\mathrm{wr}}(t) \equiv F_{\mathrm{wr}}(t)/F_{\mathrm{wr},0}$ is the 
weathering rate, $f_A(t) \equiv A_c(t)/A_{c,0}$ is the continental area, and
$f_{\mathrm{sr}}(t) \equiv S(t)/S_0$ is the areal spreading rate, which are all
normalized by their present values of Earth.  
Eq.~(\ref{gfr}) can be rearranged by introducing the geophysical forcing ratio
GFR \citep{volk87} as
\begin{equation}
f_{\mathrm{wr}}(T_{\mathrm{surf}},P_{\mathrm{CO}_2})=\frac{f_{\mathrm{sr}}}{f_A} \ =: \ \mathrm{GFR}(t)
.
\label{gfr2}
\end{equation}
Here we assume that the weathering rate only depends on the global surface temperature
and the atmospheric CO$_2$ concentration.
For the investigation of a super-Earth under external forcing,
we adopt a model planet with a prescribed continental area.  The fraction of
continental area with respect to the total planetary surface $f_A$ is varied between
$0.1$ and $0.9$.

\begin{table*}
\caption{Size and gravity parameters of the Gl~581g models}             
\label{param}      
\centering                          

\begin{tabular}{l l l l l l}        
\hline\hline                 
Parameter & \multicolumn{3}{c}{Value}   & Unit & Description \\    
 ...      & $1 M_\oplus$ & $3.1 M_\oplus$ & $4.3 M_\oplus$ & ...  & ...         \\
\hline                        
$g$         & {\y1}9.81       & {\x1}16.5              & {\x1}19.2               & m s$^{-2}$  & gravitational acceleration  \\
$R_p$       &   6378          &    8657                &    9456                 &   m         & planetary radius            \\
$R_c$       &   3471          &    4711                &    5146                 &   m         & inner radius of the mantle  \\
$R_m$       &   6271          &    8511                &    9298                 &   m         & outer radius of the mantle  \\
\hline                                   
\end{tabular}
\end{table*}

The connection between the stellar parameters and the planetary climate can be
obtained by using a radiation balance equation \citep{williams98}
\begin{equation}
\frac{L}{4\pi R^2} [1- a (T_{\mathrm{surf}}, P_{\mathrm{CO}_2})]
 = 4I_R (T_{\mathrm{surf}}, P_{\mathrm{CO}_2}),
\label{L}
\end{equation}
where $a$ denotes the planetary albedo, $I_R$ the outgoing infrared flux, and $R$ the distance from the central star.
The climate model does not include clouds, which are particularly important for determining the inner
boundary of the HZ \citep{selsis07}.
The Eqs.~(\ref{gfr2}) and (\ref{L}) constitute a set of two coupled equations with two unknowns,
$T_{\mathrm{surf}}$ and $P_{\mathrm{CO}_2}$, if the parameterization of the weathering rate, the luminosity, the
distance to the central star and the geophysical forcing ratio are specified.  Therefore, a numerical solution can be
attained in a straightforward manner.

The photosynthesis-sustaining HZ around Gl~581 is defined as the spatial domain of all distances $R$ from
the central star where the biological productivity is greater than zero, i.e.,
\begin{equation}
{\mathrm{pHZ}} \ := \ \{ R \mid \Pi (P_{\mathrm{CO}_2}(R,t), T_{\mathrm{surf}}(R,t))>0 \}.
\label{hz}
\end{equation}
In our model, biological productivity is considered to be solely a function of
the surface temperature and the CO$_2$ partial pressure in the atmosphere.
Our parameterization yields 
zero productivity for $T_{\mathrm{surf}} \leq 0^{\circ}$C or $T_{\mathrm{surf}}
\geq 100^{\circ}$C or $P_{\mathrm{CO}_2}\leq 10^{-5}$ bar \citep{franck00a}.
The inner and outer boundaries of the pHZ do not depend on
the detailed parameterization of the biological productivity within the temperature
and pressure tolerance window.

\subsection{Comments on the thermal evolution model}

Parameterized convection models are the simplest models for investigating the thermal
evolution of terrestrial planets and satellites. They have been successfully applied to the 
evolution of Mercury, Venus, Earth, Mars, and the Moon \citep{stevenson83,sleep00}.
\cite{franck95} studied the thermal and volatile history of Earth and Venus in the
framework of comparative planetology. The internal structure of massive terrestrial planets
with one to ten Earth masses has been investigated by \cite{valencia06} to
obtain scaling laws for total radius, mantle thickness, core size, and average density as 
a function of mass. Similar scaling laws were found for different compositions. We will
use these scaling laws for the mass-dependent properties of super-Earths and also the
mass-independent material properties given by \cite{franck95}.

The thermal history and future of a super-Earth has to be determined to
calculate the spreading rate for solving key Eq.~(\ref{gfr}).
A parameterized model of whole mantle convection including the volatile exchange
between the mantle and surface reservoirs \citep{franck95,franck98} is applied.
The key equations used in our present study are in accord with our previous work
focused on Gl~581c and Gl~581d; see \cite{vonbloh07}, for details.
A key element is the computation of the areal spreading rate $S$; note that $S$
is a function of the average mantle temperature $T_m$, the 
surface temperature $T_{\mathrm{surf}}$, the heat flow from the mantle $q_m$, and the
area of ocean basins $A_0$ \citep{turcotte02}.  It is given as
\begin{equation}  S = \frac{q_m^2 \pi
\kappa A_0}{4 k^2 (T_m - T_\mathrm{surf})^2}\,, 
\end{equation}
where $\kappa$ is the thermal diffusivity and $k$ the thermal conductivity.
To calculate the spreading rate, the thermal evolution of the mantle has be to computed:
\begin{equation} {4 \over 3} \pi \rho c (R_m^3-R_c^3) \frac{dT_m}{dt} = -4 \pi
R_m^2 q_m + {4 \over 3} \pi E(t) (R_m^3-R_c^3), \label{therm} \end{equation}
where $\rho$ is the density, $c$ is the specific heat at constant pressure,
$E$ is the energy production rate by
decay of radiogenic heat sources in the mantle per unit volume, and $R_m$ and $R_c$ are the
outer and inner radii of the mantle, respectively.
To calculate the thermal evolution for a planet with several Earth masses, i.e.,
3.1 and 4.3~$M_\oplus$ as pursued for Gl~581g in our present study, the planetary
parameters have to be adjusted.  Thus we assume
\begin{equation}
\frac{R_p}{R_{\oplus}}= \left(\frac{M}{M_{\oplus}}\right)^{0.27}~,
\end{equation}
where $R_p$ is the planetary radius, see \cite{valencia06}.
The total radius, mantle thickness, core size, and average density are all functions
of mass, with the subscript $\oplus$ denoting Earth values.
The exponent of $0.27$ has been obtained for super-Earths.  The values
of $R_p$, $R_m$, $R_c$, as well as the other planetary properties are scaled
accordingly.  It means that $R_p$, $R_m$ and $R_c$ increase by a factor of
1.36 for $M = 3.1 M_\oplus$ and 1.48 for $M = 4.3 M_\oplus$.

Table~\ref{param} gives a summary of the size parameters for the models
of the planets with 3.1 and 4.3~$M_\oplus$; see the study by \cite{vonbloh07}
for additional information.  The values for an Earth-size planet are included
for comparison.   The onset of plate tectonics on massive terrestrial
planets is a topic of controversy.  While \cite{oneill07} stated that there
might be in an episodic or stagnant lid regime, \cite{valencia09} proposed
that a more massive planet is likely to convect in a plate
tectonic regime similar to Earth. Thus, the more massive the planet is, the higher the
Rayleigh number that controls convection, the thinner the top boundary layer (lithosphere),
and the higher the convective velocities.  In the framework of our model,
a plate-tectonic-driven carbon cycle is considered necessary for carbon-based life.
This approach follows the previous work by \cite{vonbloh07}, who gave a detailed
discussion of the equations and parameters used in their study for the super-Earths
with 5 and 8~$M_\oplus$, identified as Gl~581c and Gl~581d, respectively.

\begin{figure}[h]
\centering
\resizebox{0.87 \hsize}{!}{\includegraphics{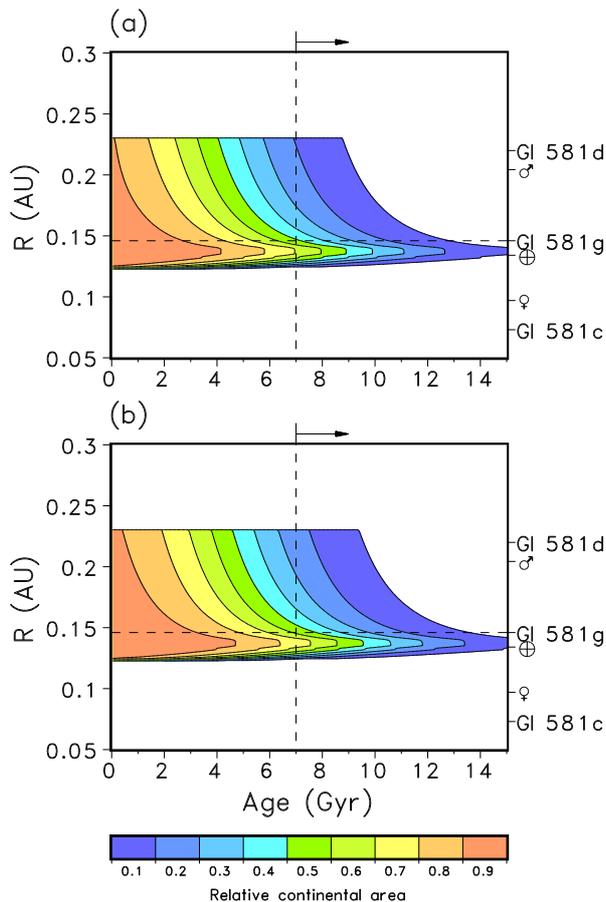}}
\caption{The pHZ of Gl~581 for a super-Earth of (a) $M=3.1 M_\oplus$ and (b) $M=4.3 M_\oplus$
as a function of planetary age.  The relative continental area is varied from 0.1 to 0.9.
The stellar luminosity is assumed to be $0.013 L_\odot$.  Following \cite{selsis07} and
references therein, the stellar age is at least 7~Gyr, indicated by a dashed line.
For comparison, we show the positions of Venus, Earth, and Mars scaled to the luminosity of Gl~581.
Note that the position of Gl~581g almost exactly coincides with the luminosity-scaled
position of Earth.
}
\label{fig2}
\end{figure}

\begin{figure}[h]
\centering
\resizebox{0.82 \hsize}{!}{\includegraphics{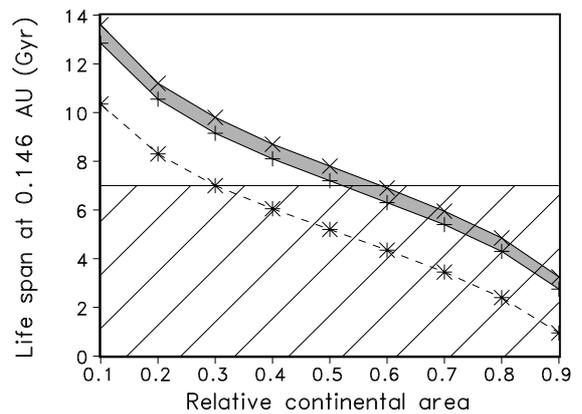}}
\caption{Life span for Gl~581g with $M=3.1 M_\oplus$ ($+$) and $M=4.3 M_\oplus$ ($\times$)
as a function of the relative continental area.  For comparison the dashed curve denotes the
life span of an Earth-mass planet. The horizontal line indicates the lower limit for
the age estimate of 7 Gyr.}
\label{fig3}
\end{figure}

\section{Results and discussion}

The photosynthesis-sustaining habitable zone is calculated for super-Earth planets with
3.1 and 4.3~$M_\odot$ and the results are depicted in Fig.~\ref{fig2}. The maximum life span
of a biosphere is given at the point in time when the pHZ vanishes.  This maximum life span
strongly depends on the relative continental area $r$ and increases with decreasing $r$.
Therefore,  ``water worlds" are favored in the facilitation of habitability as
previously obtained in models of fictitious super-Earth planets for 47~UMa and
55~Cnc \citep{franck03,vonbloh03}.  In this context, water worlds are planets
of non-vanishing continental area mostly covered by oceans.  The climate of a planet
fully covered by oceans is not stabilized by the carbonate--silicate cycle.
The maximum life span for a water world with
$r=0.1$ and a planetary mass of $M=3.1 M_\oplus$ is $15.4$~Gyr, whereas it is $16.3$~Gyr
for $M=4.3 M_\oplus$.  Both times are considerably longer than the estimated age limit of Gl~581,
which is 7~Gyr \citep{selsis07}.

This maximum life span can be realized for a super-Earth at a distance from the central star
of about $0.14 \pm 0.015$~AU.  This uncertainty is reflective of the uncertainty in the
luminosity estimate of Gl~581, noting however that our models are also subject to various
systematic uncertainties (see Sect.~2).  A full assessment of those uncertainties, which
are expected to be larger than the impact of the uncertainty in the stellar luminosity,
can best be obtained by a comparison with future alternative models.
It is noteworthy that the orbital position of Gl~581g at $0.146$~AU, if existing,
is largely consistent with this ideal position.  Thus, we can conclude that Gl~581g has
an almost perfect Goldilock position in the star--planet system of Gl~581, and it is the best
candidate for extra-terrestrial habitability so far. Because of the Goldilock position, located
well within the HZ, clouds play no major role for our results.  Any shift (within limits) of the
inner or outer boundary of the HZ will not affect the maximum life span obtained by our model.
The optimum position is not only the geometric mean of the pHZ, but it is also the position
where the maximum life span of the biosphere is realized.

But the pHZ also strongly depends on the planetary age.  Figure~\ref{fig3} depicts
the life span for a super-Earth with $3.1$ and $4.3$ $M_\oplus$ at $0.146$~AU as a function
of the relative continental area $r$.  For a relative continental area larger than $0.6$, the realized
life span is shorter than 7~Gyr, which is the estimated lower limit for the stellar age of Gl~581
star--planet system.  In this case, no habitable conditions on Gl~581g would exist
in the framework of the adopted geodynamic model.
If we place an Earth twin ($r \simeq 0.29$) at the orbital position of Gl~581g, habitable conditions
would cease just at an age of 7~Gyr. In conclusion, we have to await future missions to
identify the pertinent geodynamical features of Gl~581g (in case it does exist) and to search
for biosignatures in its atmosphere to gain insight into whether or not Gl~581g harbors life.


\end{document}